\documentclass[a4paper]{article}
\usepackage[comma,numbers,square,sort&compress]{natbib}
\usepackage{graphicx,amsmath,amssymb}
\usepackage{amsthm,bm}
\theoremstyle{definition} %% plain, definition, remark
\usepackage{subfigure,pgf,tikz,pst-node}
\usetikzlibrary{arrows,shapes,automata,positioning,fit}
\usepackage{bm} %%向量的斜体加粗
\newcommand{\lm}{\lambda}
\newcommand{\e}{\varepsilon}
\newcommand{\R}{\mathbb{R}}

\newtheorem{thmj}{Theorem}

\newtheorem{lemj}{Lemma}
\newtheorem{assj}{Assumption}

\newtheorem{remj}{Remark}

\newcommand{\pb}{\noindent\textbf{Proof. } }
\newcommand{\pe}{\hfill\rule{4pt}{8pt}}

\def\rm{\mathrm}
\renewcommand{\top}{{\mathrm{T}}}

\begin{document}
\title{Optimal Consensus for Uncertain High-order Multi-agent Systems by Output Feedback
	\thanks{This work was supported by National Natural Science Foundation of China under Grants 61973043.}
}
\author{Yutao Tang and Kui Zhu
	\thanks{Y. Tang and K. Zhu are both with the School of Artificial Intelligence, Beijing University of Posts and Telecommunications, Beijing 100876, China (e-mails: yttang@bupt.edu.cn, kuizhu\_19@butp.edu.cn). }
}	 
		
\date{}
	
\maketitle
	
{\noindent\bf Abstract}:  The distributed optimal output consensus problem for high-order multi-agent systems has been studied recently. In this paper, we further focus on the same problem for high-order multi-agent systems subject to parametric uncertainties and aim at  distributed robust controllers by measurement output feedback. We first develop a dynamic compensator to estimate the expected optimal consensus point and convert the problem into several decentralized robust tracking problems. Then, by combining the integral control technique and dirty derivative observer technique, we constructively propose a distributed output feedback integral controller to solve this problem under a mild graph connectivity condition.


\section{Introduction}

Consensus problem has been studied for decades with many applications in different areas including information fusion in sensor networks, formation control of multiple robotics,  resource allocation in power systems and so on. Basically, consensus indicates that agents will agree on a common value regarding a certain quantity of interest through repeatedly communicating with each other according to a prescribed communication pattern. In some practical applications \cite{johansson2008decentralized,nedic2010constrained, boyd2011distributed, thunberg2016optimal, wu2019optimal, yang2017distributed}, we may expect the consensus point to  enjoy an extra optimality  besides this elementary requirement . 

Recently, optimal consensus has attracted many researchers from different fields \cite{shi2013reaching, lou2014approximate, nedic2018distributed, yang2017distributed}. Typically, we suppose each agent is equipped with a convex cost function and try to construct distributed rules such that all agents reach a consensus about the optimal solution minimizing an aggregated cost function defined as the sum of their own cost functions. Various effective algorithms have been delivered in the literature from the mathematical programming viewpoint to solve this problem \cite{yuan2015gradient,kia2015distributed, yang2016PI,zeng2017distributed, li2018distributed,li2019distributed}. %Remarkably, solving these distributed optimization problems is somehow equivalent to achieving an optimal consensus for these integrators while the consensus point is specified as the optimal point of the global objective function.   

Along with the above results for single-integrator  agents, optimal (output) consensus for multi-agent systems of high-order dynamics has also been paid attention to.  In fact, engineered multi-agent systems are hardly modeled by single integrators and often with high-order physical dynamics, e.g. the robotics in radio source seeking problems \cite{kim2014cooperative}. Note that directly using the above algorithms might fail to ensure an optimal consensus for these high-order multi-agent systems. Therefore, it is crucial to design effective protocols for agents having non-integrator dynamics to complete the optimal output consensus goal. In fact, some interesting attempts have been made for both second and high-order linear agents in the literature \cite{zhang2017distributed, xie2017global,qiu2019distributed, zhao2017distributed, tang2019cyb}. However, most of these protocols are derived based on the knowledge of exact system matrices of the agents. %of agent's system matrices.  

In many practical problems,  the system matrices of each agent might be only known with measurement and modeling errors or even unavailable. This leads us to further study the solvability of  optimal output consensus when multi-agent systems are high-order with uncertain parameters.  Regarding the global consensus requirement for agents of nontrivial dynamics and also uncertain parameters, achieving an optimal consensus can be very difficult. In fact, the previous controllers in existing works \cite{xie2017global,zhang2017distributed,zhao2017distributed,tang2019cyb} can not be used any more in these circumstances and the related optimal output consensus problem should be resolved in a more aggressive manner to admit uncertain parameters in system matrices. %In other words,  we need to ensure the expected optimal consensus for multi-agent systems parameterized by the uncertainties.

To handle the parametric uncertainties, there are at least two standard strategies in the literature, i.e., adaptive control policy with a dynamic compensator and robust control policy assuming these parameters are bounded. Both strategies have been utilized to tackle the considered optimal consensus problem for several classes of multi-agent systems having uncertain parameters \cite{wang2016distributed,tang2018ijrnc,tang2021optimal,tang2021lcss,wang2021distributed}. However, most existing results require the full state of each agent except the work for relative-degree-one agents \cite{wang2016distributed}. Although this work has been extended to the case when agents are in normal form, it still requires the partial states for feedback \cite{tang2021optimal}.  Note that the state information may not be available in practice due to physical constraints and high measurement costs  \cite{teel1995tools, kim2011output, tong2017observer, tang2020distributed}. Thus,  the fundamental problem is how to constructively develop robust output feedback controllers  for the uncertain high-order agents such that the expected optimal consensus can be reached irrespective of  parametric uncertainties.

In this paper, we focus on the optimal output consensus problem for multi-agent systems having general linear dynamics with uncertain parameters. Moreover, we assume that these uncertain parameters are contained in a known compact set.  Our main goal is to seek distributed robust output feedback controllers for these agents to reach an optimal output consensus under weight-balanced directed graphs. Different from existing optimal consensus designs, we have to develop a fixed-gain output feedback controller such that an optimal consensus is robustly achieved over these parameterized linear agents, which is certainly much more challenging than conventional optimal consensus results for known linear agents by state feedback. To our knowledge, such a robust optimal output consensus  problem under the same conditions has not been addressed yet and its solvability is still unclear. 

Based on the aforementioned observations, we summarize the contributions of this paper as follows.  
\begin{itemize}
	\item First, we formulate and solve a robust optimal consensus problem for uncertain linear multi-agent systems. Different from many existing optimal consensus results, we require  the optimal consensus to be achieved irrespective of parametric uncertainties, which is definitely more challenging.  To solve the problem, we develop a novel distributed controller for the agents by integral control technique. Thus, our design has removed the requirement of a prior knowledge of the agent's system matrices and thus significantly extended some existing results \cite{xie2017global, zhao2017distributed,qiu2019distributed, tang2019cyb}  on this topic to a larger class of uncertain high-order agents.
	\item Second, we solve the optimal output consensus  problem for this class of uncertain multi-agent systems only using the measurement output information. Compared with the state/output feedback protocols in existing works \cite{zhang2017distributed, zhao2017distributed, xie2017global,qiu2019distributed, tang2018ijrnc,tang2019cyb}, this work takes both parametric uncertainties and output feedback issues into consideration. Novel distributed controllers are constructively developed by combining the integral control technique and dirty derivative high-gain observer technique, which may save measurement cost and communication resource.
	\item Third, by taking some special quadratic cost functions for the agents, we provide an alternative way to reach an average consensus for the agents.  In contrast to some existing publications for integrator agents \cite{ren2008distributed, rezaee2015average},  the average output consensus problem can be solved in a robust manner for more general high-order agents by only measurement output feedback.
\end{itemize}

The rest of this paper is organized as follows. We first present some preliminaries and the problem statement in Sections \ref{sec:pre} and \ref{sec:form}. Then we detail the main results in Section \ref{sec:solvability} with proofs. Numerical examples are provided to verify the effectiveness of our designs in Section \ref{sec:simu} with some closing remarks in Section \ref{sec:con}.

\section{Preliminaries}\label{sec:pre}

In this paper, $\R^N$ stands for the $N$-dimensional Euclidean space. Let $\mbox{col}(a_1,\,\dots,\,a_N)=[a_1\,\dots\, a_N]^\top$ and $I_{N}$ be the identity matrix of size $N$.  Let $||a||$ be the Euclidean norm of a vector $a$ and  $||A||$ the spectral norm of a matrix $A$.  %The subscript might be omitted when it is self-evident. 

\subsection{Graph theory}

A weighted directed graph (digraph) is given by $\mathcal {G}=(\mathcal {N},\, \mathcal {E},\, \mathcal{A})$ with node set $\mathcal{N}$, edge set $\mathcal {E}\in \mathcal{N} \times \mathcal{N}$, and weighted adjacency matrix $\mathcal{A}$. We avoid self-loops in the graph and define  $a_{ii}=0$ and $a_{ij}\geq 0$ if there exists an edge $(j,\,i)\in \mathcal{E}$. We say this digraph is strongly connected if there is a directed path for any two vertices.

Define the in-degree and out-degree of node $i$ as  $d^{\mbox{in}}_i=\sum\nolimits_{j=1}^N a_{ij}$ and $d^{\mbox{out}}_i=\sum\nolimits_{j=1}^N a_{ji}$. A digraph is weight-balanced if the in-degree and out-degree are equal for any node. The Laplacian  of  $\mathcal{G}$ is given as $L\triangleq D^{\mbox{in}}-\mathcal{A}$ with $D^{\mbox{in}}=\mbox{diag}(d^{\mbox{in}}_1,\,\dots,\,d^{\mbox{in}}_N)$. For a weight-balanced and strongly connected digraph, the eigenvalues of matrix $\mbox{Sym}(L)\triangleq \frac{L+L^\top}{2}$ are positive real numbers and can be ordered as $0=\lambda_1<\lambda_2\leq \dots\leq \lambda_N$.   More details can be found in  \cite{godsil2001algebraic}.

\subsection{Convex analysis}

Consider a function $f\colon \R^m \rightarrow \R$. We say it is convex if for any $0\leq a \leq 1$ and $\zeta_1,\zeta_2 \in \R^m$, we have  $f(a\zeta_1+(1-a)\zeta_2)\leq af(\zeta_1)+(1-a)f(\zeta_2)$. When this function $f$ is differentiable, we denote by $\nabla f$ its gradient. It can be verified that for a differentiable function $f$, it is convex if and only if for  $\forall \zeta_1,\zeta_2 \in \mathbb{R}^m$, we have $f(\zeta_1)-f(\zeta_2)\geq \nabla f(\zeta_2)^\top (\zeta_1 -\zeta_2)$. If this inequality holds strictly except the trivial case, we say $f$ is strictly convex. A function $f$ is $\omega$-strongly convex ($\omega >0$) over $\R^m$ if $(\nabla f(\zeta_1)-\nabla f(\zeta_2))^\top (\zeta_1 -\zeta_2)\geq \omega \|\zeta_1 -\zeta_2\|^2$ holds for for any  $\zeta_1, \zeta_2 \in \R^m$. Consider a function ${\bm f}\colon \R^m \rightarrow \R^m$.  If $\|{\bm f}(\zeta_1)-{\bm f}(\zeta_2)\|\leq \vartheta \|\zeta_1-\zeta_2\|$ holds for any $ \zeta_1, \zeta_2 \in \R^m$,  we say  it is $\vartheta$-Lipschitz . Please refer to the monograph \cite{bertsekas2003convex} for more details if interested.
%\end{align*}

\section{Problem statement} \label{sec:form}

Consider an $N$-agent system of the following form: 
\begin{align}\label{sys:agent}
	\begin{split}
		\dot{{x}}_i& = A(w){x}_i + B(w) {u}_i, \\
		y_i&=C(w)  {x}_i, \quad i=1,\,\ldots, \,N
	\end{split}
\end{align}
with state variable ${x}_i \in \R^{n}$, input variable $u_i\in \R$, and output variable  $y_i \in \mathbb{R}$. Here  $A(w)\in \R^{n\times n}$, $B(w)\in \R^{n\times 1}$, and $C(w)\in \R^{1\times n}$ are the system matrices subject to an uncertain parameter vector $w\in \mathcal{W} \subset \R^{n_w}$. Without loss of generality, the nominal value is set to be $w={\bm 0}$. We also assume $w$ can be any vector ranging over a compact set $\mathcal{W}$ containing the origin.

We equip each agent $i$ with a convex cost function $f_i\colon \R \to \R$. Let $f(y)=\sum\nolimits_{i=1}^{N} f_i(y)$ be the global cost function with $y$ the decision variable.  This paper aims at an output consensus for these agents about the minimal solution of  $f$ in a distributed manner irrespective of any $w\in \mathcal{W}$. 

We use a digraph $\mathcal{G}=(\mathcal{N},\, \mathcal{E}, \,\mathcal{A})$ to represent the communication network for these agents with $\mathcal{N}=\{1,\,\dots,\, N\}$. If agent $i$ can receive the information from agent $j$, then there is a weighted edge $(j,\,i)\in \mathcal{E}$ in $\mathcal{G}$ with $a_{ij}>0$. Denote by $\mathcal{N}_i=\{j \mid (j,\, i)\in \mathcal {E}\}$ the neighbor set of node $i$ for each node $i\in \mathcal{N}$.  

Regarding the multi-agent system \eqref{sys:agent} with any given compact set $\mathcal{W}$, cost function $f_i(\cdot)$, and a digraph $\mathcal{G}$, the robust distributed optimization output consensus problem or simply robust DOOC problem is to derive a measurement output feedback controller $u_i$ using agent $i$'s local information such that, for any $w\in \mathcal{W}$,  the agent trajectories from any initial points are well-defined for any $t\geq 0$  and the agent outputs achieve an optimal output consensus in the sense that  $\lim_{t\to +\infty}||y_i(t)-y^*||=0$ with $y^*$ being an optimal solution to the following global optimization problem
\begin{align}\label{opt:main}
	\min_{ y \in \R} \; f(y)=\sum\nolimits_{i=1}^{N} f_i(y)
\end{align}

\begin{remj}\label{rem:formulation}
	This robust DOOC formulation extends the existing optimal consensus results in the literature to the case when agents are of high-order dynamics possessing parametric  uncertainties. Furthermore, we aim at distributed controllers by output feedback instead of the widely used state feedback to handle the optimality and robustness issues.  Thus, our problem is  more general and more challenging than the well-studied (optimal) consensus for integrators  \cite{rezaee2015average, xie2017global,qiu2019distributed,tang2019cyb}.  
\end{remj}

In a special case, we can choose a constant $c_i>0$ and let $f_i(y)=c_i(y-y_i(0))^2$ for each $i \in \mathcal{N}$. Then the optimal consensus point will be a weighted average of agents' initial outputs. Hence, the formulation will assist us in resolving the weighted average consensus problem for uncertain agent \eqref{sys:agent} by output feedback. 

As optimal consensus for non-integrator agents is nontrivial when the communication graph is directed, the limited measurements and parametric uncertainties bring us extra technical difficulties. In next section, we will adopt the embedded control approach proposed in \cite{tang2019cyb} and extend it to a robust version  to tackle these issues.

\section{Main Results} \label{sec:solvability}

We will first convert the optimal output consensus  problem into some decentralized tracking problems with the help of an optimal signal generator to estimate the expected global optimal point, and then complete the whole design by solving the resultant robust tracking problems by output feedback controllers.

To ensure the solvability of the robust DOOC problem for agent \eqref{sys:agent}, we make the following assumptions.

\begin{assj}\label{ass:convexity-strong}
	For each $i \in \mathcal{N}$, there exist constants $\underline{l}_i,\, \overline l_i>0$ such that $f_i$ is $\underline{l}_i$-strongly convex and   $\nabla f_i$  is $\overline l_i$-Lipschitz.
\end{assj}

\begin{assj}\label{ass:graph}
	The digraph $\mathcal{G}$ is strongly connected and weight-balanced.
\end{assj}

\begin{assj}\label{ass:relative-degree}
	Agent \eqref{sys:agent} has a well-defined relative degree $m$  and is minimum-phase for all $w\in \mathcal{W}$.
\end{assj}

These assumptions are very standard and have been widely used in the literature \cite{kia2015distributed, wang2016distributed, kim2011output, tang2019cyb}. Under Assumption \ref{ass:convexity-strong},  the optimization problem \eqref{opt:main} must have a unique optimal solution.  We denote it by $y^*$ and assume it is finite as usual \cite{kia2015distributed,zhang2017distributed}.   Under Assumption \ref{ass:graph}, each agent's information can finally reach another agent in this network.   Assumption \ref{ass:relative-degree} ensures that the high-frequency gain $b_1(w)=C(w)A^{m-1}B(w)$ does not vanish for any $w\in \mathcal{W}$. We assume that $b_1(w)>0$ for all $w\in \mathcal{W}$ without loss of generality. This assumption  can characterize  a large class of uncertain high-order agents, including integrators and exactly known linear systems \cite{ren2008distributed, rezaee2015average, xie2017global, zhao2017distributed, tang2019cyb} as some special cases.

\subsection{Optimal signal generation}

Following the embedded design procedure, we will first consider a single-integrator multi-agent system specified by $\dot{z}_i=\mu_i$ with the function $f_i$ and digraph $\mathcal{G}$. After solving the optimal consensus for them, we will convert our optimal output consensus problem into several robust output tracking problems for agent \eqref{sys:agent}  with reference $z_i$. 

Since the information graph  $\mathcal{G}$ is directed, its Laplacian  might be asymmetric. Thus, the optimal signal generator proposed in \cite{tang2019cyb} fails to achieve our goal without the information of $L^\top$. To tackle this issue, the following optimal signal generator has been developed in \cite{tang2021optimal} for problem \eqref{opt:main} under weighted-balanced graphs:
\begin{align}\label{sys:generator}
	\begin{split}
		\dot{z}_i&=-\alpha \nabla f_i(z_i)-\beta \sum\nolimits_{j=1}^{N}a_{ij}(z_i-z_j)+\sum\nolimits_{j=1}^{N}a_{ij}(v_i-v_j)\\
		\dot{v}_i&=\alpha \beta  \sum\nolimits_{j=1}^{N}a_{ij}(z_i-z_j)
	\end{split}
\end{align}
with some chosen parameters $\alpha,\,\beta>0$.  Its effectiveness has already been established in \cite{tang2021optimal}. Here we denote $\underline{l}=\min_i\{l_i\}$, $\bar l=\max_i\{\bar l_i\}$ and present a sketch of proof for a complete design. 

\begin{lemj}\label{lem:generator}
	Under Assumptions \ref{ass:convexity-strong}--\ref{ass:graph}, we let  
	\begin{align}\label{eq:parameter:alpha-beta}
		\alpha\geq \max\{1,\,\frac{1}{\underline{l}},\,\frac{2\bar l^2}{\underline{l}\lm_2}\},\quad \beta\geq \max\{1,\, \frac{1}{\lambda_2},\,\frac{6\alpha^2\lambda_N^2}{\lambda_2^2} \}
	\end{align} 
	Then, for any $z_i(0)$ and $v_i(0)$, the trajectory $z_i(t)$ under the algorithm \eqref{sys:generator} will exponentially converge to the global optimal point $y^*$ as $t\to \infty$ for any $i\in \mathcal{N}$, i.e., for all $t>0$, $||z_i(t)-y^*||\leq c_1 e^ {-c_2 t}$ holds for two constants $c_1, c_2>0$.
\end{lemj}
\pb   Letting $z=\mbox{col}(z_1,\,\dots,\,z_N)$ and $v=\mbox{col}(v_1,\,\dots,\,v_N)$, we can put \eqref{sys:generator} into a compact form
\begin{align}\label{sys:composite-osg}
	\dot{z}&=-\alpha \nabla \tilde f(z)- \beta Lz-Lv,\quad \dot{v}=\alpha \beta Lz
\end{align}
where $\tilde f(z)\triangleq \sum\nolimits_{i=1}^Nf_i(z_i)$. It can be verified that $\tilde f(z)$ is $\underline{l}$-strongly convex while its gradient $\nabla\tilde f(r)$ is $\bar l$-Lipschitz.  Assuming $\mbox{col}(z^{\star},\,v^{\star})$ be any equilibrium point of system \eqref{sys:composite-osg}, one can easily verify $z^{\star}={\bm 1}_N y^{\star}$ under Assumptions \ref{ass:convexity-strong}--\ref{ass:graph}. 

Then, we let  $M_1=\frac{1}{\sqrt{N}} {\bm 1}_N$ and $M_2$ be the matrix satisfying $M_2^\top M_1={\bm 0}_{N-1}$, $M_2^\top M_2=I_{N-1}$ and $M_2 M_2^\top=I_{N}-M_1 M_1^\top$. Set $\bar z_1=M_1^\top (z-z^\star)$, $\bar z_2=M_2^\top (z-z^\star)$,  and $\bar v_2=M_2^\top[( v+\alpha z)-( v^\star+\alpha z^\star)]$. It follows that  
\begin{align}\label{sys:composite-osg-reduced}
	\begin{split}
		\dot{\bar z}_1&=-\alpha M_1^\top {\bm \Pi}\\
		\dot{\bar z}_2&=-\alpha M_2^\top {\bm \Pi}-\beta M_L \bar z_2 + \alpha M_L \bar z_2-M_L\bar v_2\\
		\dot{\bar v}_2&=-\alpha M_L {\bar v}_2+\alpha^2 M_L \bar z_2-\alpha^2 M_2^\top {\bm \Pi} 
	\end{split}
\end{align}
where  ${\bm \Pi}\triangleq \nabla \tilde f(z)-\nabla \tilde f(z^\star)$ and $M_L= M_2^\top LM_2$.  Choose a Lyapunov function candidate as $W_{\rm o}(\bar z_1,\,\bar z_2,\, \bar v_2)=\frac{1}{2}||\bar z_1||^2+\frac{1}{2}||\bar z_2||^2+\frac{1}{2\alpha^3}||\bar v_2||^2$. It is quadratic and positive definite. By a similar treatment as in  \cite{tang2021optimal}, the derivative of $W_{\rm o}(t)$ along the trajectory of \eqref{sys:composite-osg-reduced} satisfies 
$
\dot{W}_{\rm o} \leq -\frac{1}{2}W_{\rm o}
$.  According to Theorem 4.10 in   \cite{khalil2002nonlinear},  $W_{\rm o}(\bar z_1(t),\,\bar z_2(t),\,\bar v_2(t))$ will exponentially converge to $0$ as $t$ goes to $\infty$. Since $z-{\bm 1}_N y^{\star}=M_1 \bar z_1+M_2\bar z_2$, the proof is complete.
\pe

This optimal signal generator is motivated by the primal-dual method solving a distributed optimization problem \cite{gharesifard2014distributed, kia2015distributed}.  Compared with the ones in \cite{kia2015distributed} and  \cite{zhang2017distributed}, this algorithm is free of initialization via exchanging $v_i$ with each other.  With the generator \eqref{sys:generator}, agent $i$ will get an asymptotic estimate $z_i(t)$ of $y^*$. Thus, we are going to construct effective output tracking controllers for each agent such that the local tracking error $y_i(t)-z_i(t)$ asymptotically vanishes.  If done, by inserting  \eqref{sys:generator} into the developed robust tracking controllers, we will have a distributed controller solving our problem for agent \eqref{sys:agent}.

When the parameter $w$ is exactly known by us, the resultant output tracking problem for agent \eqref{sys:agent} can be readily solved by some well-known stabilization-based techniques, e.g., pole placement and linear quadratic regulation \cite{chen1995linear}. However, these unknown parameters  make the design of such stabilization-based controllers very tricky. To make it clear,  we recall Proposition 9.1.1 in  \cite{isidori1995nonlinear} and put agent \eqref{sys:agent} into a normal form by some coordinate transformation: 
\begin{align}\label{sys:normal-form}
	\begin{split}
		\dot{x}^0_i&=A_0(w)x_{i0}+b_0(w)y_i\\
		\dot{\xi}_{ir}&={\xi}_{ir+1},\quad r=1,\dots,\,m-1\\
		\dot{\xi}_{im}&=A_1(w)x_{i0}+A_2(w)\xi_i+b_1(w)u_i\\
		y_i&=\xi_{i1}
	\end{split}
\end{align}
where $\xi=\mbox{col}(\xi_{i1},\,\dots,\,\xi_{im})$, $\xi_{ir}=y^{(r-1)}_{i}$ for $r=1,\dots,\,m$, and $A_0$, $A_1$, $A_2$, $b_0$, $b_1$ are (possible unknown) constant matrices with proper dimensions. The symbol $y^{(r)}_{i}$ represents the $r$-th derivative of $y_i(t)$ with respect to $t$. It can be found that the pole-placement method in  \cite{zhang2017distributed} and  \cite{tang2019cyb} highly relies on the knowledge of matrices $A_1(w)$, $A_2(w)$, and $b_1(w)$, which is prohibitive in our setting. Technically,  we have to stabilize several parameterized matrices at the same time by a fixed output feedback controller, which is certainly more difficult than the conventional pole placement problem for known linear systems by state feedback. 

Some adaptive and/or robust control policies  have been utilized in the literature \cite{wang2016distributed,tang2018ijrnc,zhang2017distributed,tang2021optimal} to deal with such an uncertain parameter issue for an optimal output consensus. However, when facing agents of the form \eqref{sys:agent} (or equivalently, \eqref{sys:normal-form}), they all require the agent's full state information (or at least partial state information) and can not be directly implemented to solve the formulated robust DOOC problem for the multi-agent system  \eqref{sys:agent} when only measurement output information is available for us. 

To overcome these issues, we are going to explore the well-known integral control and high-gain observer techniques to design robust tracking controllers for agent \eqref{sys:normal-form}. In the sequel, we will first establish an important lemma to ensure an optimal consensus by a partial state feedback control, and then extend it to an output feedback version to solve our problem.  

\subsection{Partial state feedback integral control}

Suppose the variables $y_i$, $\dot{y}_i$, $\dots$, $y_i^{(m-1)}$ are all known to us.  We choose positive constants $k_1$, $\dots$, $k_m$ such that the polynomial $p(s)\triangleq k_1+k_2s+\dots+k_ms^{m-1}+s^{m}$ is Hurwitz.  For simplicity, we can set
\begin{align}\label{eq:parameter:k}
	k_j=\binom{m}{j-1}\lambda_0^{m-j-1},\quad j=1,\,\dots,\,m
\end{align} 
with any given $\lambda_0>0$. Then the polynomial can be written as $p(s)=(s+\lambda_0)^m$ and is apparently Hurwitz. 

To deal with the parametric uncertainties, we denote $e_{vi}=y_i-z_i$ and introduce a compensator as the integral term: 
\begin{align}
	\dot{\xi}_{i0}=e_{vi}
\end{align}
Then a distributed partial state feedback integral controller can be given for system \eqref{sys:normal-form} as follows:
\begin{align}\label{ctr:part-state}
	u_i&=-\e[k_1\xi_{i0}+k_2(y_i-z_i)+k_3\dot{y}_i+\dots+k_m  {y}^{(m-2)}_i+y_i^{(m-1)}]\nonumber\\ 
	\dot{\xi}_{i0}&=y_i-z_i\nonumber\\ 
	\dot{z}_i&=-\alpha \nabla f_i(z_i)-\beta \sum\nolimits_{j=1}^{N}a_{ij}(z_i-z_j)+\sum\nolimits_{j=1}^{N}a_{ij}(v_i-v_j)\nonumber\\ 
	\dot{v}_i&=\alpha \beta  \sum\nolimits_{j=1}^{N}a_{ij}(z_i-z_j)
\end{align}
with constants $k_1 ,\,\dots,\,k_m$, $\alpha$,\, $\beta$ chosen as above and gain parameter $\e>0$ to be specified later. 

To establish the effectiveness of our controller \eqref{ctr:part-state}, we  let $U(w)=-b_1^{-1}(w)[A_1(w)A_0^{-1}(w) b_0(w)+A_2(w)E]$ and $E=\mbox{diag}(1,\,0,\,\dots,\,0)$ and  introduce some new variables:  
\begin{align*}
	\bar{x}^0_i&=x_{i0}+A_0^{-1}(w) b_0(w)z_i\\
	\bar \xi_{i0}&=\xi_{i0}+\frac{U(w)z_i}{\e k_1}\\
	\bar \xi_{i1}&=e_{vi}, \quad \bar \xi_{ir}=\xi_{ir}, \quad \bar \xi_i=\mbox{col}(\bar \xi_{i1},\,\dots,\,\bar \xi_{im}), \quad r=2,\dots,m\\
	\sigma_i&=\sum\nolimits_{j=1}^{m}k_j\bar \xi_{ij-1}+\bar \xi_{im}\\
	{\bar \xi}_{ie}&=\mbox{col}(\bar \xi_{i0},\,\dots,\,\bar \xi_{im-1}) 
\end{align*}
By some standard mathematical manipulations, we can derive the following translated tracking error system for agent \eqref{sys:normal-form} with an output reference $z_i(t)$: 
%\begin{align*}%\label{sys:normal-form}
%\begin{split}
%\dot{\bar x}_{i0}&=A_0(w)\bar x_{i0}+b_0(w)\bar e_{vi}+A_0^{-1}(w)b_0(w)\dot{z}_i\\
%\dot{\bar \xi}_{i0}&=\bar \xi_{i1}+\frac{U(w)}{\e k_1}\dot{z}_i\\
%\dot{\bar \xi}_{i1}&={\bar \xi}_{i2}-\dot{z}_i\\
%\dot{\bar \xi}_{ir}&={\bar \xi}_{ir+1},\quad r=2,\dots,\,m-1\\
%\dot{\bar \xi}_{im}&=A_1(w)\bar x_{i0}+A_2(w)\bar \xi_i+b_1(w)[u_i-U(w)z_i]\\
%y_i&=\xi_{i1}
%\end{split}
%\end{align*}
%
%Letting $\sigma_i=\sum\nolimits_{j=1}^{m}k_j\bar \xi_{ij-1}+\bar \xi_{im}$, system \eqref{sys:normal-form} is further transformed into the following form.
\begin{align}\label{sys:translated}
	\begin{split}
		\dot{\bar x}_{i0}&=A_0(w)\bar x_{i0}+b_0(w)\bar \xi_{i1}+D_x(w)\dot{z}_i\\
		\dot{\bar \xi}_{ie}&=\bar A_0 {\bar \xi}_{ie}+ \bar b_0 \sigma_i+D_\xi (w)\dot{z}_i \\
		\dot{\sigma}_{i}&=A_1(w)\bar x_{i0}+\bar A_2(w){\bar \xi}_{ie}+\bar A_3(w) \sigma_i+D_{\sigma}(w)\dot{z}_i +b_1(w)[u_i-U(w)z_i]
	\end{split}
\end{align}
where matrices $\bar A_0$, $\bar b_0$, $D_\xi$, $\bar A_2$, $\bar A_3$, $D_x$, $D_\sigma$ are defined as follows.
\begin{align*}
	\bar A_0&=\left[\begin{array}{c|c}
		{\bm 0}_{m-1}&I_{m-1} \\ \hline
		-k_1&[-k_2,\,\dots,\,-k_m] 
	\end{array}\right],\quad 
	\bar b_0=\begin{bmatrix} 
		{\bm 0}_{m-1}\\ 1
	\end{bmatrix},\quad 
	D_\xi (w)=\begin{bmatrix}
		\frac{U(w)}{\e k_1}\\
		-1\\
		{\bm 0}_{m-2}
	\end{bmatrix}\\
	\bar A_2(w)&=A_2(w)\bar A_0-k_m[k_1,~k_2-\frac{k_1}{k_m},\, \dots,\, k_m-\frac{k_{m-1}}{k_m}],\quad \bar A_3(w)=A_2(w)\bar b_0+k_m\\
	D_x(w)&= A_0^{-1}(w) b_0(w), \quad D_\sigma(w)=\frac{U(w)}{\e}-k_2
\end{align*}

Substituting the controller \eqref{ctr:part-state} into system \eqref{sys:normal-form} gives:
\begin{align}\label{sys:whole:part-state}
	\dot{\bar x}_{i0}&=A_0(w)\bar x_{i0}+b_0(w)\bar \xi_{i1}+D_x(w)\dot{z}_i  \nonumber \\
	\dot{\bar \xi}_{ie}&=\bar A_0 {\bar \xi}_{ie}+ \bar b_0 \sigma_i+D_\xi (w)\dot{z}_i \nonumber \\
	\dot{\sigma}_{i}&=A_1(w)\bar x_{i0}+\bar A_2(w){\bar \xi}_{ie}+[\bar A_3(w)-\e b_1(w)]\sigma_i+D_\sigma(w)\dot{z}_i \nonumber  \\
	\dot{z}_i&=-\alpha \nabla f_i(z_i)-\beta \sum\nolimits_{j=1}^{N}a_{ij}(z_i-z_j)+\sum\nolimits_{j=1}^{N}a_{ij}(v_i-v_j) \nonumber \\
	\dot{v}_i&=\alpha \beta  \sum\nolimits_{j=1}^{N}a_{ij}(z_i-z_j)
\end{align}

Denote $\bar x_i\triangleq \mbox{col}({\bar x}_{i0}, {\bar \xi}_{ie}, {\sigma}_{i})$. We give a lemma to ensure the solvability of our optimal output consensus problem by \eqref{ctr:part-state}. 
\begin{lemj}\label{thm:part-state}
	Suppose Assumptions \ref{ass:convexity-strong}--\ref{ass:relative-degree} hold. The robust DOOC problem for multi-agent system \eqref{sys:agent} with \eqref{opt:main} can be  solved by the partial state feedback controller  \eqref{ctr:part-state} with some chosen constants $k_1,\,\dots,\,k_m$, $\alpha$, $\beta$, and $\e$.
\end{lemj}
\pb  With $k_1 ,\,\dots,\,k_m$, $\alpha$, $\beta$ chosen as above, we only have to determine an $\e$ such that the expected optimal consensus is achieved for agent \eqref{sys:normal-form} under  \eqref{ctr:part-state}.  

The following proof consists of two steps.  

{\em Step 1}: we show that the $\bar x_i$-subsystem of \eqref{sys:whole:part-state} is input-to-state stable with respect to input $\dot{z}_i$ for a large enough constant $\e$. %To prove this, we use Lyapunov arguments and seek a Lyapunov function. 
Since matrices $A_0(w)$ and $\bar A_0$ are Hurwitz by the choices of $k_1,\,\dots,\,k_m$ under Assumption \ref{ass:relative-degree}, there must exist two positive definite matrices $P_0$ and $P_1$ uniquely for any fixed $w\in \mathcal{W}$ such that  $A_0^\top (w) P_0(w)+P_0(w)A_0(w)=-2I_{n-m}$ and $A_1^\top P_1+P_1A_1=-2I_{m} $ are both satisfied.

Let $V_i(\bar x_i)=\bar x_{i0}^\top P_0(w){\bar x_{i0}}+\hat\epsilon{\bar \xi}_{ie}^\top P_1 {\bar \xi}_{ie}+\sigma_i^2$ with $\hat\epsilon>0$ to be given later. We take its time derivative along the trajectory of  \eqref{sys:whole:part-state} and obtain that
\begin{align*}
	\dot{V}_i & = 2\bar x_{i0}^\top P_0(w)[A_0(w)\bar x_{i0}+b_0(w)\bar \xi_{i1}+D_x(w)\dot{z}_i] +2\hat\epsilon{\bar \xi}_{ie}^\top P_1 [\bar A_0 {\bar \xi}_{ie}+ \bar b_0 \sigma_i+D_\xi (w)\dot{z}_i]+2\sigma_iA_1(w)\bar x_{i0}\\
	&+2\sigma_i\{\bar A_2(w){\bar \xi}_{ie}+[\bar A_3(w)-\e b_1(w)]\sigma_i +D_\sigma(w) \dot{z}_i\}\\
	&=-2\bar x_{i0}^\top \bar x_{i0}+2\bar x_{i0}^\top P_0(w)b_0(w)\bar \xi_{i1}+2\bar x_{i0}^\top P_0(w)D_x(w)\dot{z}_i-2\hat\epsilon{\bar \xi}_{ie}^\top {\bar \xi}_{ie} +2\hat\epsilon{\bar \xi}_{ie}^\top P_1  \bar b_0 \sigma_i+2\hat\epsilon{\bar \xi}_{ie}^\top P_1 D_\xi (w)\dot{z}_i\\
	&+2\sigma_i A_1(w)\bar x_{i0}+2[\bar A_3(w)-\e b_1(w)]\sigma_i^2+2\sigma_i\bar A_2(w){\bar \xi}_{ie}+2\sigma_i D_\sigma(w)\dot{z}_i
\end{align*}
Using Young's inequality to dominate the cross terms at the righthand side, we have that
\begin{align*}
	\dot{V}_i &\leq -||\bar x_{i0}||^2+ 2||P_0(w)b_0(w)||^2||\bar \xi_{i1}||^2+ 2||P_0(w)D_x(w)||^2 ||\dot{z}_i||^2 -\hat\epsilon||{\bar \xi}_{ie}||^2 + 2\hat\epsilon ||P_1  \bar b_0||^2 ||\sigma_i||^2\\
	& + 2\hat\epsilon ||P_1 D_\xi (w)||^2 ||\dot{z}_i||^2 +2 [\bar A_3(w)-\e b_1(w)]\sigma_i^2+ 2 ||A_1(w)||^2\sigma_i^2 + \frac{1}{2}||\bar x_{i0}||^2 +  \frac{2}{\hat\epsilon}||\bar A_2(w)||^2 \sigma_i^2+\frac{\hat\epsilon}{2}||{\bar \xi}_{ie}||^2\\
	&+ \sigma_i^2+||D_\sigma(w)||^2||\dot{z}_i||^2\\
	&\leq -\frac{1}{2}||\bar x_{i0}||^2 -(\frac{\hat\epsilon}{2}-2||P_0(w)b_0(w)||^2)||{\bar \xi}_{ie}||^2 - 2 [\e b_1(w) -\hat\epsilon ||P_1  \bar b_0||^2-\Xi_{i \sigma}(w)-1]\sigma_i^2 +\Xi_{iz}(w)||\dot{z}_i||^2 
\end{align*}
where $\Xi_{i\sigma}(w) \triangleq \bar A_3(w)-\hat\epsilon ||P_1  \bar b_0||^2-||A_1(w)||^2-\frac{1}{\hat\epsilon}||\bar A_2(w)||^2$ and $\Xi_{iz}(w)\triangleq 2[||P_0(w)D_x(w)||^2 +\hat\epsilon ||P_1 D_\xi (w)||^2 +||D_\sigma(w)||^2] $.

Choosing 
\begin{align}\label{eq:parameter:epsilon}
	\begin{split}
		\hat\epsilon\geq 4\max_{w\in \mathcal{W}}\{||P_0(w)b_0(w)||^2\}+1,\quad \e\geq \frac{\max_{w\in \mathcal{W}}\Xi_{i \sigma}(w)+\hat\epsilon ||P_1  \bar b_0||^2+2}{\min_{w \in W}\{b_1(w)\}}
	\end{split}
\end{align}
gives  $\dot{V}_i \leq -\frac{1}{2}||\bar x_{i0}||^2 - ||{\bar \xi}_{ie}||^2 - 2 \sigma_i^2+\Xi_{iz}(w)||\dot{z}_i||^2$.  This implies the following for some constants $\bar c_1,\, \bar c_2>0$ that 
\begin{align}\label{eq:iss-part-state}
	\dot{V}_i &\leq -\bar c_1V_i+\bar c_2 ||\dot{z}_i||^2
\end{align}

{\em Step 2}: we show the solvability of our robust DOOC problem using inequality \eqref{eq:iss-part-state}. Under Assumption \ref{ass:convexity-strong}, the function $\dot{z}_i$ is globally Lipschitz at $\mbox{col}(\bar z,\, \bar v)$. At the same time, $z_i(t)$ and $\dot{z}_i(t)$ exponentially converge to $y^*$ and $0$ as $t\to \infty$ by Lemma \ref{lem:generator}, respectively. Then we can solve the differential inequality \eqref{eq:iss-part-state} and obtain 
\begin{align*}
	{V}_i(t) &\leq e^{ -\bar c_1(t-t_0)}V_i(t_0)+ \bar c_2 \int_{t_0}^t e^{-\bar c_1(t-\tau)}||\dot{z}_i(\tau)||^2{\rm d}\tau
\end{align*}
From this, one concludes that  $V_i(t)$, $\bar x_i(t)$, and $\bar \xi_{i1}(t)$ all exponentially converge to $0$ as $t\to\infty$. Using the fact that $|y_i-y^*|\leq |y_i-z_i|+|z_i-y^*|$,   we obtain the exponential convergence of $y_i(t)$ towards $y^*$ as $t\to\infty$. The proof is thus complete.
\pe

It is interesting to point out that the proposed controller \eqref{ctr:part-state} requires less information than the one used in \cite{tang2019cyb} to solve the DOOC problem. In fact, we take the zero dynamics $x_i^0$ as dynamic uncertainties and compensate them by high-gain controls here. In this way, the controller \eqref{ctr:part-state} has a lower order than those proposed in \cite{tang2019cyb}. 

Next, motivated by the observer-based designs in previous works \cite{kim2011output,tong2017observer,tang2020distributed}, we move on to develop an output feedback controller extending the partial-state feedback \eqref{ctr:part-state} by  high-gain dirty observers and thus complete the whole robust DOOC design. %for multi-agent system \eqref{sys:agent}.

\subsection{Output feedback integral control}

To this end, we propose a dirty derivative observer to estimate these unknown derivatives as in  \cite{teel1995tools}: 
\begin{align}\label{sys:dirty}
	\begin{split}
		\dot{\chi}_{ir}&={\chi}_{ir+1}-l_r(\chi_{i1}-y_i),\quad r=1,\dots,\,m-1\\
		\dot{\chi}_{im}&=-l_m(\chi_{i1}-y_i)
	\end{split}
\end{align}
Here ${\chi}_{i}\triangleq \mbox{col}({\chi}_{i1},\,\dots,\,{\chi}_{im})$ and $l_r=\gamma^r k_{m-r+1}$ with a constant $\gamma>0$ to be specified later.  

Substituting these estimations into the partial-state feedback control \eqref{ctr:part-state}, we present an output feedback control to solve our robust DOOC problem as follows.
\begin{align}\label{ctr:output}
	u_i&=-\e[k_1\xi_{i0}+k_2(y_i-z_i)+k_3\chi_{i2}+\dots+k_m \chi_{im-1}+\chi_{im}]  \nonumber\\
	\dot{\xi}_{i0}&=y_i-z_i \nonumber\\
	\dot{\chi}_{ir}&={\chi}_{ir+1}-l_r(\chi_{i1}-y_i),\quad r=1,\dots,\,m-1  \nonumber\\
	\dot{\chi}_{im}&=-l_m(\chi_{i1}-y_i)  \nonumber\\
	\dot{z}_i&=-\alpha \nabla f_i(z_i)-\beta \sum\nolimits_{j=1}^{N}a_{ij}(z_i-z_j)+\sum\nolimits_{j=1}^{N}a_{ij}(v_i-v_j) \nonumber\\
	\dot{v}_i&=\alpha \beta  \sum\nolimits_{j=1}^{N}a_{ij}(z_i-z_j)
\end{align}
where constants $k_1 ,\,\dots,\,k_m$, $\alpha$, $\beta$ and $\e$ are chosen as in Lemma \ref{thm:part-state}. This controller for agent $i$ only depends its local information and thus is distributed.

Here is our main theorem to solve the optimal output consensus problem for uncertain multi-agent system \eqref{sys:agent} by distributed output feedback controllers.
\begin{thmj}\label{thm:output}
	Suppose Assumptions \ref{ass:convexity-strong}--\ref{ass:relative-degree} hold.  The robust DOOC problem for multi-agent system \eqref{sys:agent} with \eqref{opt:main} can be solved by a distributed output feedback controller of the form \eqref{ctr:output} with some chosen constants $k_1 ,\,\dots,\,k_m$, $\alpha$,  $\beta$, $\e$, and $\gamma$ .	
\end{thmj}
\pb To complete the proof, we seek for a similar inequality as \eqref{eq:iss-part-state} for some translated systems under \eqref{ctr:output}.  

Let $\bar \chi_{ir}=\chi_{ir}-\xi_{ir}$ for $r=1,\, \dots,\, m$. The error system associated with the dirty observer \eqref{sys:dirty} is:
\begin{align*}
	\begin{split}
		\dot{\bar \chi}_{ir}&=\bar {\chi}_{ir+1}-l_r\bar \chi_{i1},\quad r=1,\dots,\,m-1\\
		\dot{\bar \chi}_{im}&=-l_m\bar \chi_{i1}-A_1(w)x_{i0}-A_2(w)\xi_i-b_1(w)u_i  
	\end{split}
\end{align*}

Substituting \eqref{ctr:output} into this error system gives
\begin{align*}%\label{sys:dirty-error}
	\begin{split}
		\dot{\bar \chi}_{ir}&=\bar {\chi}_{ir+1}-l_r\bar \chi_{i1},\quad r=1,\dots,\,m-1\\
		\dot{\bar \chi}_{im}&=-l_m\bar \chi_{i1}-\Delta_i-b_1(w)\bar u_i
	\end{split}
\end{align*}
where $\Delta_i\triangleq A_1(w)\bar x_{i0}+A_2(w)\bar \xi_i-\e b_1(w)\sigma_i $ and $\bar u_i \triangleq-\e(k_3\bar \chi_{i2}+\dots+k_m \bar \chi_{im-1}+\bar \chi_{im})$.

Let $\hat \chi_{ir}=\gamma^{m-r} \bar \chi_{ir}$ for $r=1,\,\dots,\,m$. It follows that
\begin{align*}%\label{sys:dirty-error-scale}
	%\begin{split}
	\dot{\hat \chi}_{ir}&=\gamma(-k_{m-r+1}\bar \chi_{i1}+\hat {\chi}_{ir+1}),\quad r=1,\dots,\,m-1\\
	\dot{\hat \chi}_{im}&=-\gamma k_1\bar \chi_{i1}-[\Delta_i+b_1(w)\bar u_i]
	%\end{split}
\end{align*}
It has a compact form as follows. 
\begin{align*}%\label{sys:dirty-error-scale}
	%\begin{split}
	\dot{\hat \chi}_{i}&=\gamma A_{\chi}\hat \chi_i- b_{\chi}[\Delta_i+b_1(w)\bar u_i]
	%\end{split}
\end{align*}
where $A_\chi=\left[\begin{array}{c|c}
	-{\bm p}_\chi&I_{m-1} \\ \hline
	-k_1&{\bm 0}_{m-1} 
\end{array}\right]$, $b_\chi=\begin{bmatrix}
	{\bm 0}_{m-1}\\
	1
\end{bmatrix}$ with ${\bm p}_\chi=\mbox{col}(k_m,\,\cdots,\, k_2)$.  

Recalling the translated system \eqref{sys:translated}, we can use similar mathematical manipulations and obtain the tracking subsystem under controller \eqref{ctr:output}  as follows.
\begin{align}\label{sys:whole:output}
	\dot{\bar x}_{i0}&=A_0(w)\bar x_{i0}+b_0(w)\bar \xi_{i1}+D_x(w)\dot{z}_i  \nonumber\\
	\dot{\bar \xi}_{ie}&=\bar A_0 {\bar \xi}_{ie}+ \bar b_0 \sigma_i+D_\xi (w)\dot{z}_i \nonumber \\
	\dot{\sigma}_{i}&=A_1 \bar x_{i0}+\bar A_2 {\bar \xi}_{ie}+[\bar A_3 -\e b_1(w)]\sigma_i+b_1(w)\bar u_i+D_\sigma(w)\dot{z}_i \nonumber\\
	\dot{\hat \chi}_{i}&=\gamma A_{\chi}\hat \chi_i-\gamma^{1-m}b_{\chi}[\Delta_i+b_1(w)\bar u_i] 
\end{align}

From the choice of $k_i$, the equation $A_\chi^\top P_\chi+P_\chi A_\chi=-2I_{m}$ has a unique positive definite solution $P_\chi$. Let $W_i(\bar x_i,\, \hat \chi_i)=V_i(\bar x_i)+\hat \chi_i^\top P_\chi\hat \chi_i$ with $V_i$ defined as before in the proof of Lemma \ref{thm:part-state}. $W_i$ is quadratic and positive definite. 

By setting  $k_1,\,\dots,\,k_m$, and $\e$ as  in Lemma \ref{thm:part-state}, we have 
\begin{align*}
	\dot{V}_i &\leq -\frac{1}{2}||\bar x_{i0}||^2 - ||{\bar \xi}_{ie}||^2 -2 \sigma_i^2+2\sigma_i b_1(w)\bar u_i+\Xi_{iz}(w)||\dot{z}_i||^2\\
	&\leq -\frac{1}{2}||\bar x_{i0}||^2 - ||{\bar \xi}_{ie}||^2 - \sigma_i^2+ ||b_1(w)||^2||\bar u_i||^2+\Xi_{iz}(w)||\dot{z}_i||^2
\end{align*}
with $V_i$ defined as in the proof of  Lemma \ref{thm:part-state}.

Recalling the compactness of set ${\mathcal W}$, we can further determine some known constants $\hat c_1,\,\hat c_2,\,\hat c_3>0$ such that
\begin{align*}
	\dot{V}_i\leq -\hat c_1V_i+\hat c_2||\bar u_i||^2+\hat c_3 ||\dot{z}_i||^2
\end{align*} 

It follows then % The derivative of $W_i$ along the trajectory of \eqref{sys:whole:output} satisfies
\begin{align*}
	\dot{W}_i&= \dot{V}_i+2\hat \chi_i^\top P_\chi \{\gamma A_{\chi}\hat \chi_i- b_{\chi}[\Delta_i+b_1(w)\bar u_i]\}\\
	&\leq -\hat c_1V_i-2\gamma ||\hat \chi_i||^2-2 \hat \chi_i^\top P_\chi b_{\chi}[\Delta_i+b_1(w)\bar u_i]+\hat c_2||\bar u_i||^2 +\hat c_3 ||\dot{z}_i||^2 
\end{align*} 

Next, we will estimate the above cross terms by Young's inequality.  From the expression of $\Delta_i$,  there must be a known constant $l_1>0$ such that $||\Delta_i||^2\leq l_1 V_i^2$. Then 
\begin{align*}
	||2 \hat \chi_i^\top P_\chi b_{\chi}\Delta_i||&\leq \frac{2l_1}{\hat c_1}||P_\chi b_{\chi}||^2||\hat \chi_i||^2+\frac{\hat c_1}{2l_1}||\Delta_i||^2\\
	&\leq  \frac{2l_1}{\hat c_1}||P_\chi b_{\chi}||^2||\hat \chi_i||^2+\frac{\hat c_1}{2}V_i^2
\end{align*}

Note that $\bar u_i=-\e(k_3\gamma^{2-m} \hat \chi_{i2}+\dots+k_m\gamma\hat \chi_{im-1}+\hat \chi_{im})$. For any $\gamma>1$, there must be a constant $l_2>0$ satisfying $||\bar u_i||^2\leq l_2 ||\hat \chi_i||^2$. By Young's inequality, one can determine a known constant $l_3>0$ such that
\begin{align*}
	||2\hat \chi_i^\top P_\chi b_{\chi}b_1(w)\bar u_i||\leq || \bar u_i||^2+||P_\chi b_{\chi} b_1(w)||^2||\hat \chi_i||^2\leq l_3||\hat \chi_{i}||^2
\end{align*}

Putting these inequalities together, we have 
\begin{align*}
	\dot{W}_i&\leq -\hat c_1 V_i-2\gamma ||\hat \chi_i||^2+ \frac{2l_1}{\hat c_1}||P_\chi b_{\chi}||^2||\hat \chi_i||^2+\frac{\hat c_1}{2}V_i^2+l_3||\hat \chi_{i}||^2+\hat c_2||\bar u_i||^2+\hat c_3 ||\dot{z}_i||^2 \\
	&\leq  -\frac{\hat c_1}{2} V_i-(2\gamma-\hat c_2 l_2-\frac{2l_1}{\hat c_1}||P_\chi b_{\chi}||^2-l_3) ||\hat \chi_2||^2 +\hat c_3 ||\dot{z}_i||^2
\end{align*} 
Fixing the parameter $\gamma$ to satisfy
\begin{align}\label{eq:parameter:gamma}
	\gamma \geq \max \{1,\,3\hat c_2 l_2,\,\frac{6l_1}{\hat c_1}||P_\chi b_{\chi}||^2,\,3l_3\}
\end{align}
one can obtain
\begin{align*}
	\dot{W}_i&\leq -\frac{\hat c_1}{2} V_i- ||\hat \chi_i||^2+ \hat c_3 ||\dot{z}_i||^2
\end{align*}
With this inequality, we can follow the same procedure as in the proof of \eqref{thm:part-state} and conclude the solvability of the optimal output consensus problem by the output feedback control law of the form \eqref{ctr:output} under theorem conditions. 
\pe

\begin{remj}
	Compared with some similar optimal consensus results \cite{zhang2017distributed,xie2017global,zhao2017distributed,tang2019cyb}, this theorem removes the requirement of the exact system matrices.  Thanks to the output feedback integral controller \eqref{ctr:output}, the optimal output consensus problem for these  high-order agents can be solved in a robust fashion to admit arbitrarily large uncertain parameters. 
\end{remj}
\begin{remj}
	The developed controller \eqref{ctr:output} involves a number of control parameters. To determine these parameters, we are suggested to first choose $\alpha$,\,$\beta$ according to \eqref{eq:parameter:alpha-beta}, and then set $k_1,\,\dots,\,k_m$, and $\e$ according to \eqref{eq:parameter:k} and \eqref{eq:parameter:epsilon}. After that, we can increase $\gamma$ to satisfy  condition \eqref{eq:parameter:gamma}.
\end{remj}

%\begin{remj}
%	In our controllers, the parameters are mainly derived based on several norm bounds of the uncertain system matrices. One may prefer to increase $\alpha,\,\beta,\,\e, \,\gamma$ sequentially and choose some sufficient ones by repeated simulations.	
%\end{remj}

\section{Simulation}\label{sec:simu}

In this section, we provide two examples to illustrate the effectiveness of our design.

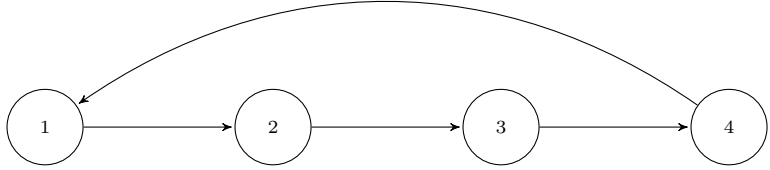
\begin{figure}
	\centering
	\begin{tikzpicture}[shorten >=1pt, node distance=3 cm, >=stealth',
		every state/.style ={circle, minimum width=1 cm, minimum height=1cm}]
		\node[align=center,state](node1) {\scriptsize 1};
		\node[align=center,state](node2)[right of=node1]{\scriptsize 2};
		\node[align=center,state](node3)[right of=node2]{\scriptsize 3};
		\node[align=center,state](node4)[right of=node3]{\scriptsize 4};
		\path[->]  (node1) edge (node2)
		(node2) edge (node3)
		(node3) edge (node4)
		(node4) edge [bend right=35]  (node1)
		;
	\end{tikzpicture}
	\caption{Digraph $\mathcal{G}$ used in our examples.} \label{fig:graph}
\end{figure}

{\it Example 1}. Consider a multi-agent system composed of four identical Vertical Takeoff and Landing (VTOL) aircrafts \cite{isidori2012robust} tasked to reach a consensus on the vertical position. The communication graph is depicted as Fig.~\ref{fig:graph} with all the edge weights as $1$. The motion of the aircraft on the lateral-vertical plane can be modeled by the equations
\begin{align*}
	M\ddot{p}_i&=-\sin(\theta_i)T_i+2\cos(\theta_i)\sin(\alpha)F_i\\
	M\ddot{q}_i&=\cos(\theta_i)T_i+2\sin(\theta_i)\cos(\alpha)F_i-gM\\
	J\ddot{\theta}_i&=2l\cos(\alpha)F_i
\end{align*}
where $p_i$, $q_i$, and $\theta_i$ denote, respectively, the horizontal and vertical position of the center of mass of the aircraft and the roll angle of the aircraft with respect to the horizon. $M$ denotes the mass of the aircraft and $J$ the moment of inertia about the center of mass. Moreover, $T_i$ is the thrust directed out from the bottom of the aircraft and $F_i$ is the equal force acting at the wingtips tilted by some fixed angle $\alpha$ to the horizontal body axis. The constants $l$ and $g$ are the distance between the wingtips and the gravitational acceleration. Since we are interested in a consensus on the vertical position, we neglect the rolling motion (i.e., $\theta_i\equiv0$) and focus on the reduced vertical dynamics described by
\begin{align*}%\label{eq:ex1}
	\ddot{q}_i&=\frac{1}{M} T_i-g
\end{align*}
The mass of the aircraft is assumed to have some constant variation satisfying $\frac{1}{2}\leq \frac{M}{M_0}\leq 2$ with a nominal mass $M_0$. To save energy, we require the consensus position to be the average of their initial positions $\mbox{Aver(z(0))}\triangleq \frac{1}{4}\sum_{i=1}^4z_i(0)$ in order to minimize the aggregate distance from their starting point to this final position. Next, we show that this task can be reformulated as an optimal output consensus problem for a group of uncertain agents of the form \eqref{sys:agent}.

Note that the agent dynamics can be rewritten as follows. 
\begin{align}\label{eq:ex1}
	\begin{bmatrix}
		\dot{q}_i\\
		\ddot{q}_i
	\end{bmatrix}
	&=\begin{bmatrix}
		0 &1\\
		0&  0
	\end{bmatrix}\begin{bmatrix}
		q_i\\
		\dot{q}_i
	\end{bmatrix}+\begin{bmatrix}
		0\\
		(1+w)\frac{1}{M_0}
	\end{bmatrix}T_i-\begin{bmatrix}
		0\\
		g
	\end{bmatrix}
\end{align}
with $w=\frac{M_0}{M}-1\in [-\frac{1}{2},\, 1]$. Letting $x_i=\mbox{col}(q_i,\,\dot{q}_i)$, $y_i=z_i$, and $u_i=T_i$, the agent is naturally of the form \eqref{sys:agent} with $m=2$ but subject to an extra actuating disturbance due to the gravity. By assigning a local cost function  $f_i(y)= ||y-z_i(0)||^2$ to each agent, the consensus task on the vertical position for these aircrafts is converted to the formulated DOOC problem for agent \eqref{eq:ex1} with a global cost function $f(z)=\sum_{i=1}^4 ||z-z_i(0)||^2$. Since Assumption \ref{ass:graph} can be confirmed with $\lambda_2=1$ and $\lambda_4=2$, this problem is readily solved by the controller \eqref{ctr:output} according to Theorem \ref{thm:output} if the external disturbance disappears.  It is interesting to point output that the designed integral controller \eqref{ctr:output} is robust enough to ensure the solvability of this robust DOOC problem even with this external disturbance due to the gravity. 

\begin{figure}
	\centering
	\includegraphics[width=0.84\textwidth]{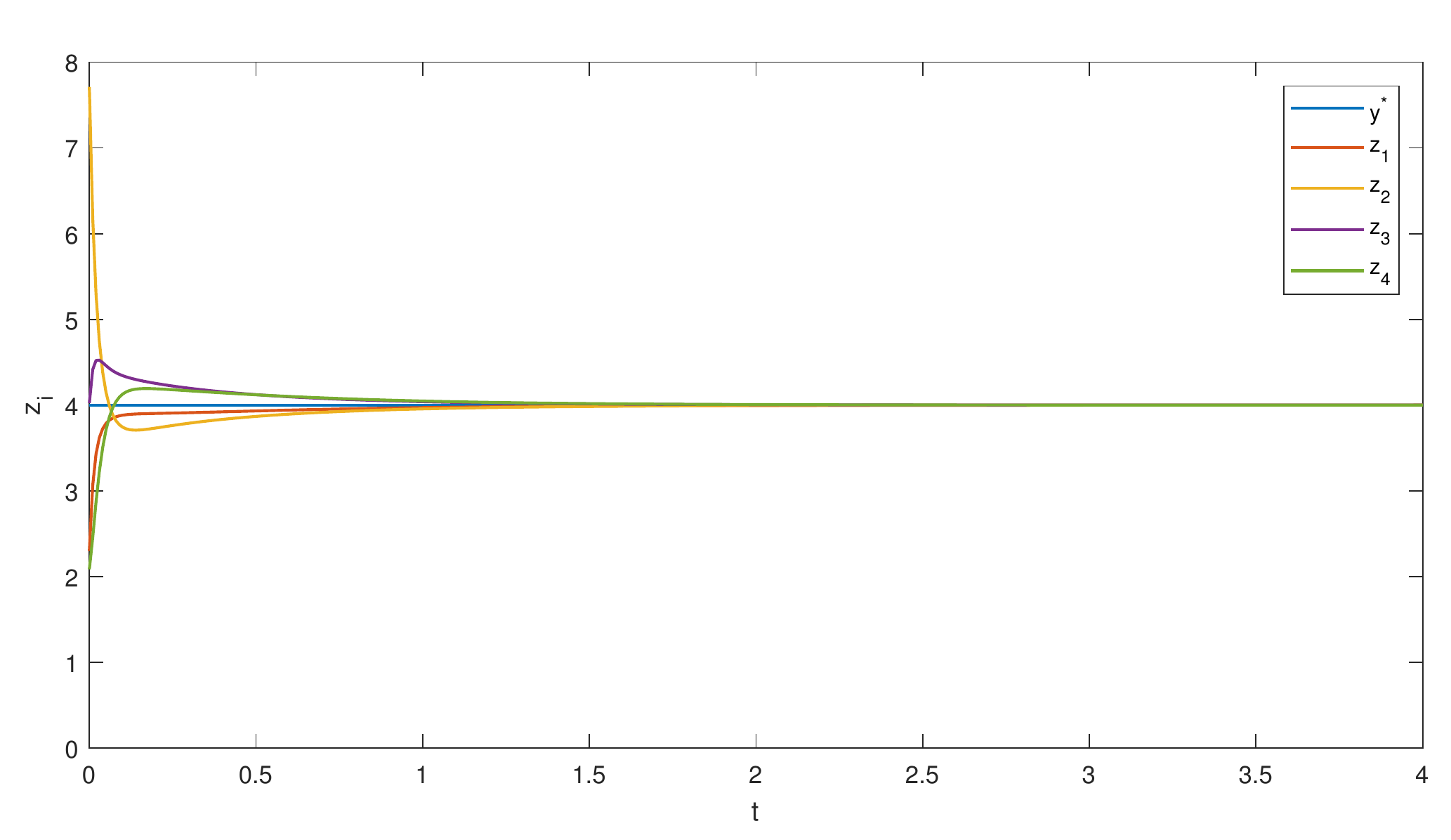}
	\caption{Estimates of the global optimal solution $y^*$ in {\it Example 1}.}	\label{fig:generator-1}
\end{figure}

\begin{figure}
	\centering
	\includegraphics[width=0.84\textwidth]{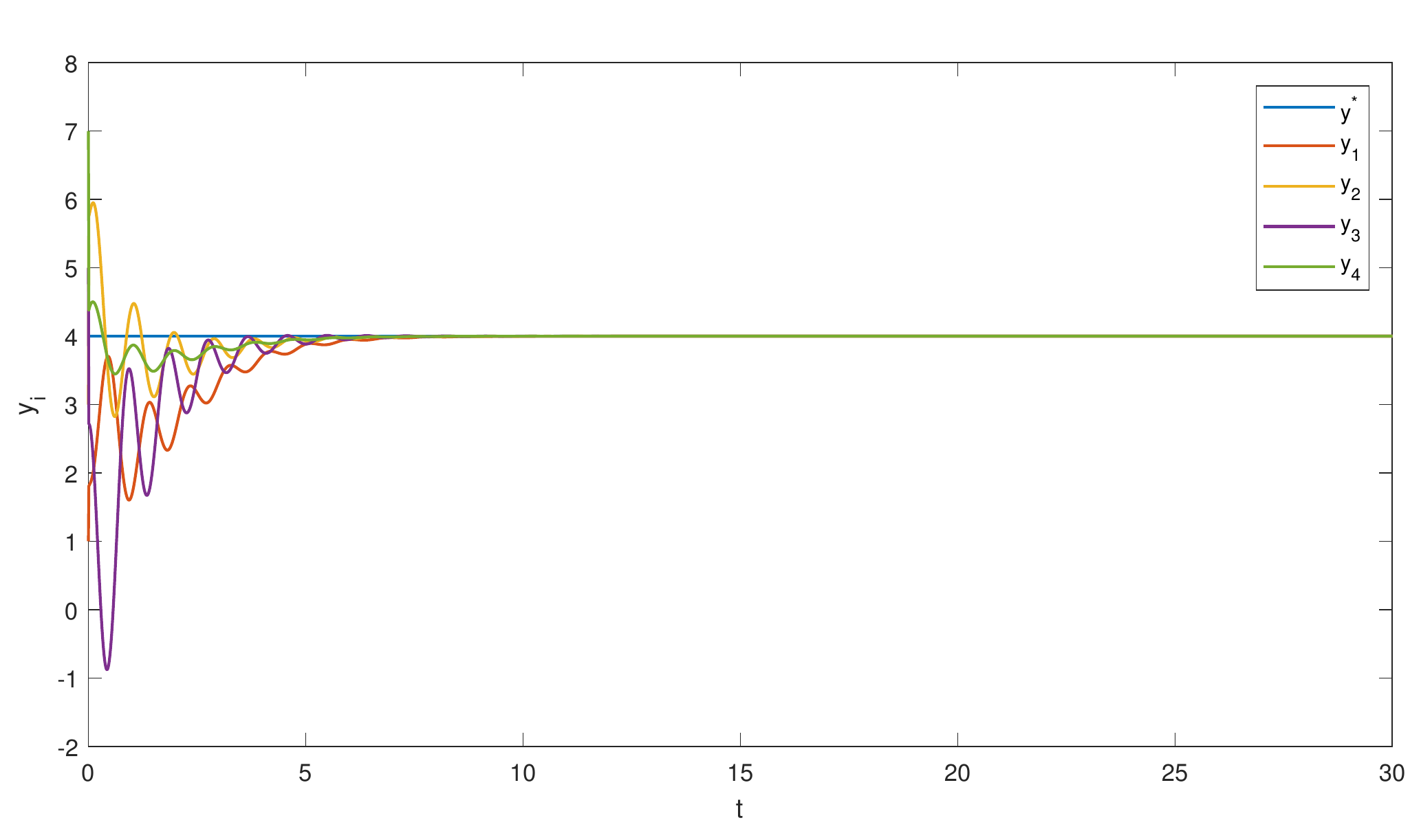}
	\caption{Profile of agent outputs under the controller \eqref{ctr:output}  in {\it Example 1}.}	\label{fig:simu1-1}
\end{figure}

\begin{figure}
	\centering
	\includegraphics[width= 0.84\textwidth]{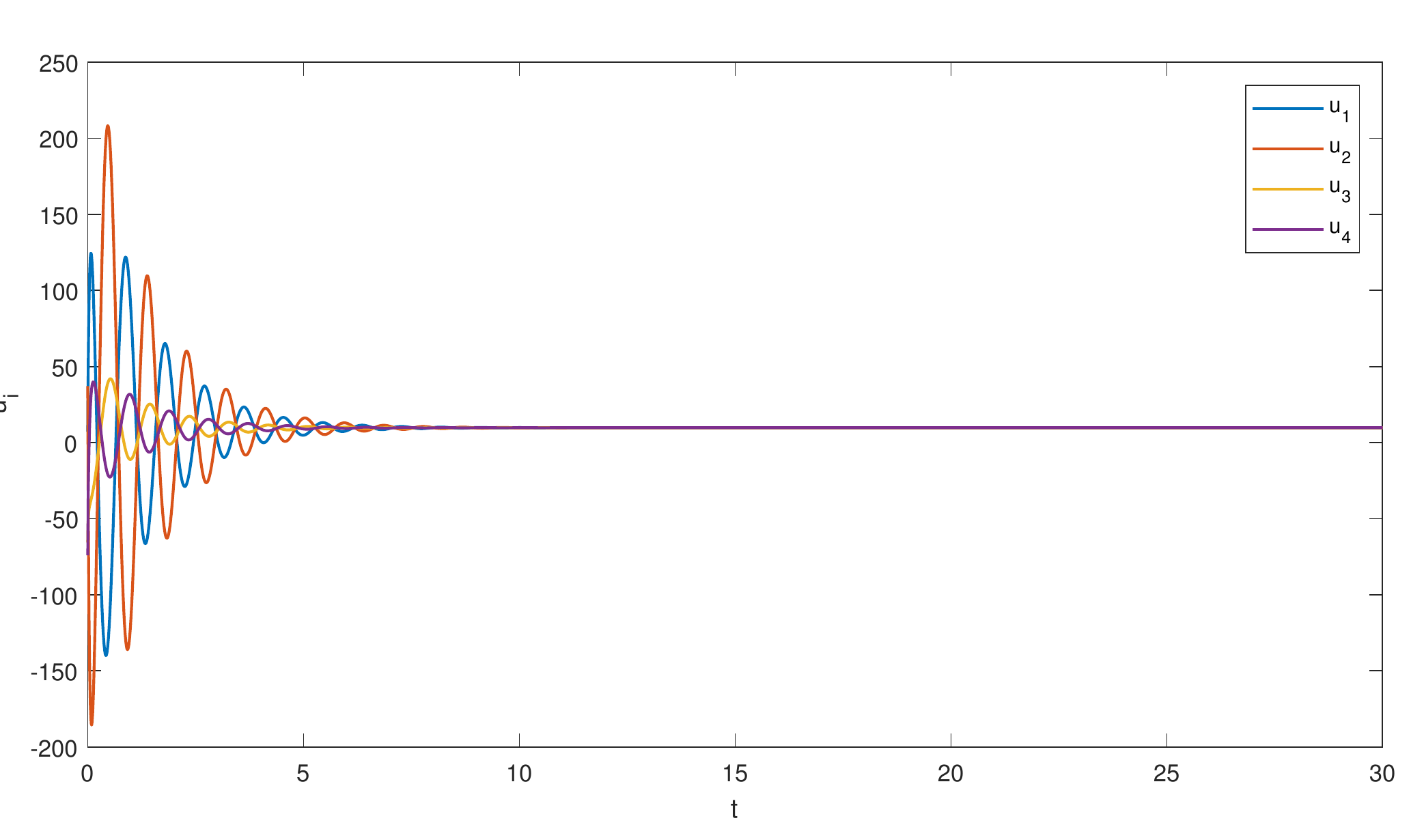}
	\caption{Profile of control efforts under the output feedback controller \eqref{ctr:output}  in {\it Example 1}. }	\label{fig:simu2-1}
\end{figure}

For simulations, we set $z_i(0)=2*i-1$ and choose $k_1=1$, $k_2=2$, $\alpha=1$, $\beta=15$, $\e=6$, and $\gamma=10$ for the controller \eqref{ctr:output}.  Other initial conditions for the closed-loop system are (randomly) chosen.  The performance of our proposed optimal signal generator \eqref{sys:generator} is illustrated in Fig.~\ref{fig:generator-1}, where the estimate $z_i$ converges to the global optimal solution $y^*=\mbox{Aver(z(0))}=4$ quickly. The expected optimal output consensus for these agents can be observed in Fig.~\ref{fig:simu1-1}, while the control signals are bounded as reported in Fig.~\ref{fig:simu2-1}.  These observations confirm the effectiveness of our controllers to solve the optimal consensus problem on the vertical position for these VTOL aircrafts.


{\it Example 2}. We further consider a four-agent system with the following system matrices and more complex cost functions.
\begin{align*}
	A(w)&=\begin{bmatrix}
		-1+w_1&1&0\\
		-1+w_2&0&1\\
		1&w_3&1
	\end{bmatrix},\quad B(w)=\begin{bmatrix}
		0\\
		0\\
		1+w_3
	\end{bmatrix},\quad C(w)=\begin{bmatrix}
		0& 1+w_4&0
	\end{bmatrix} 
\end{align*}
Here parameters $w_1,\,\dots,\,w_4$ are supposed to be unknown but satisfy  $||w_i||\leq 0.5$. Denote $w\triangleq \mbox{col}(w_1,\,w_2,\,w_3,\,w_4)$ for short. We let $\mathcal{W}=[-0.5,\,0.5]\times [-0.5,\,0.5]\times [-0.5,\,0.5]\times [-0.5,\,0.5]\subset \mathbb{R}^4$.  It can be verified that this agent has  a well-defined relative degree two and is minimum phase for all $w\in \mathcal{W}$. 

The cost functions of agents are taken as 
\begin{align*}
	{f_1}(y)=\frac{1}{2} (y-8)^2, \quad {f_2}(y) = \frac{y^2}{160\ln {({y^2} + 2} )} + \frac{1}{2}(y - 5)^2\\
	{f_3}(y)=\frac{y^2}{40\sqrt {y^2 + 1} } + \frac{1}{2}y^2, \quad {f_4}(y) =  \frac{1}{2}\ln \left( {{e^{ - 0.05{y}}} + {e^{0.05{y}}}} \right) + \frac{1}{2}y^2. 
\end{align*}
We can verify Assumption \ref{ass:convexity-strong} with $\underline{l}_i=0.5,\, \overline{l}_i=1.5$ for each $i$. The communication digraph is the same with {\it Example 1}.  According to Theorem \ref{thm:output}, the robust DOOC problem for these uncertain linear agents can be solved by a distributed output feedback controller of the form \eqref{ctr:output}.    

\begin{figure}
	\centering
	\includegraphics[width=0.84\textwidth]{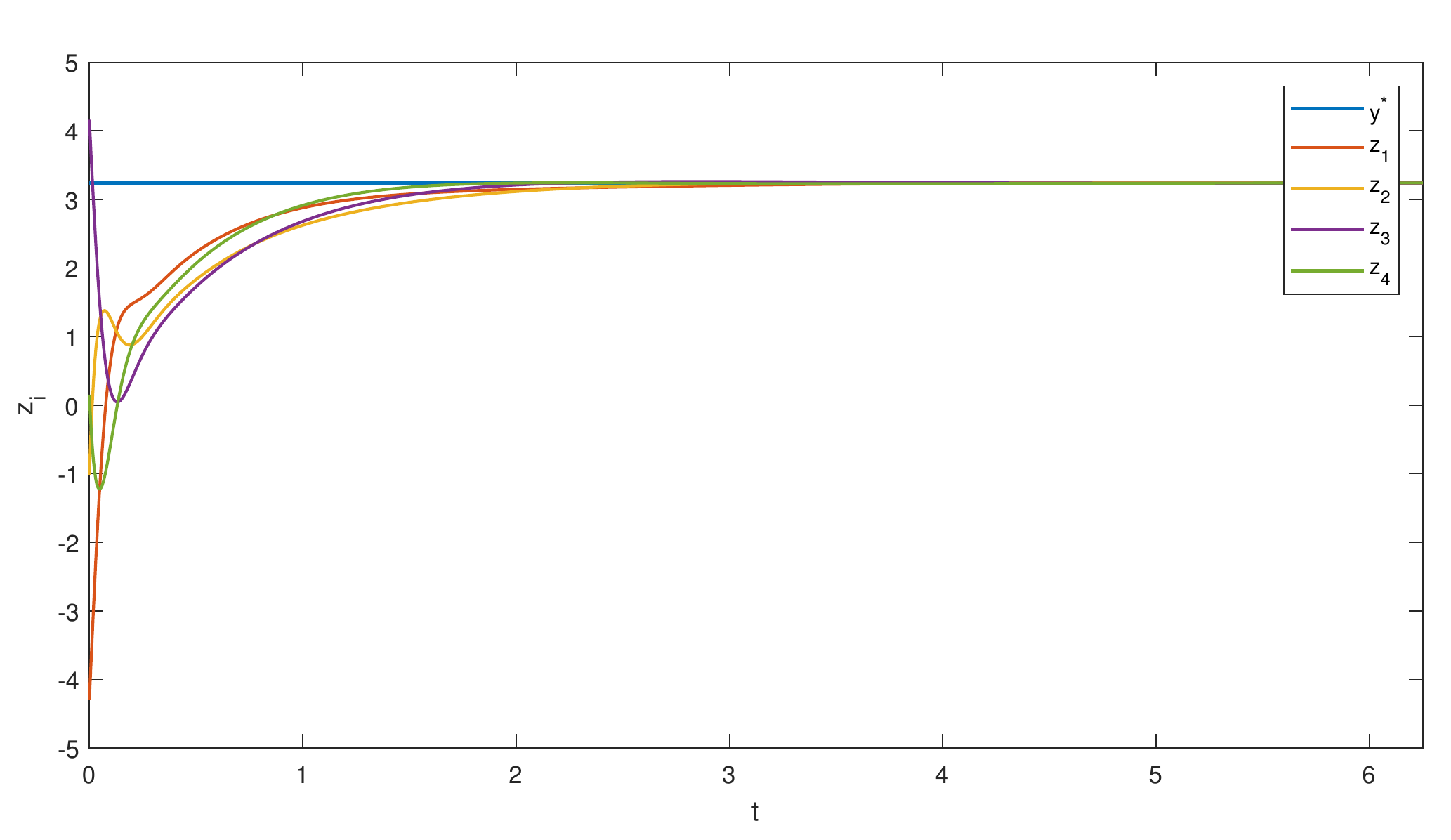}
	\caption{Estimates of the global optimal solution $y^*$   in {\it Example 2}.}	\label{fig:generator}
\end{figure}

\begin{figure}
	\centering
	\includegraphics[width=0.84 \textwidth]{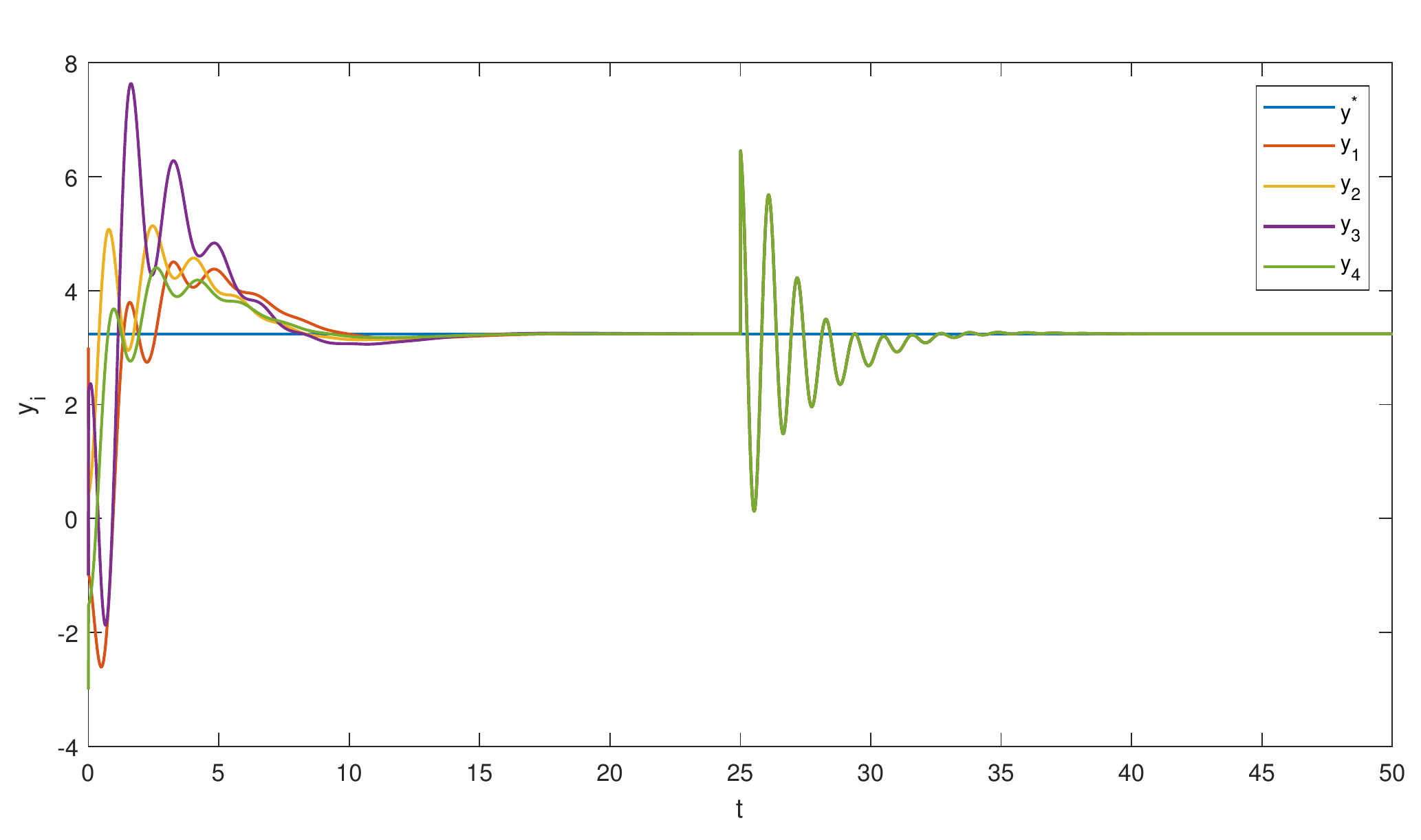}
	\caption{Profile of agent outputs under the controller \eqref{ctr:output}  in {\it Example 2}.}	\label{fig:simu1}
\end{figure}

\begin{figure}
	\centering
	\includegraphics[width= 0.84\textwidth]{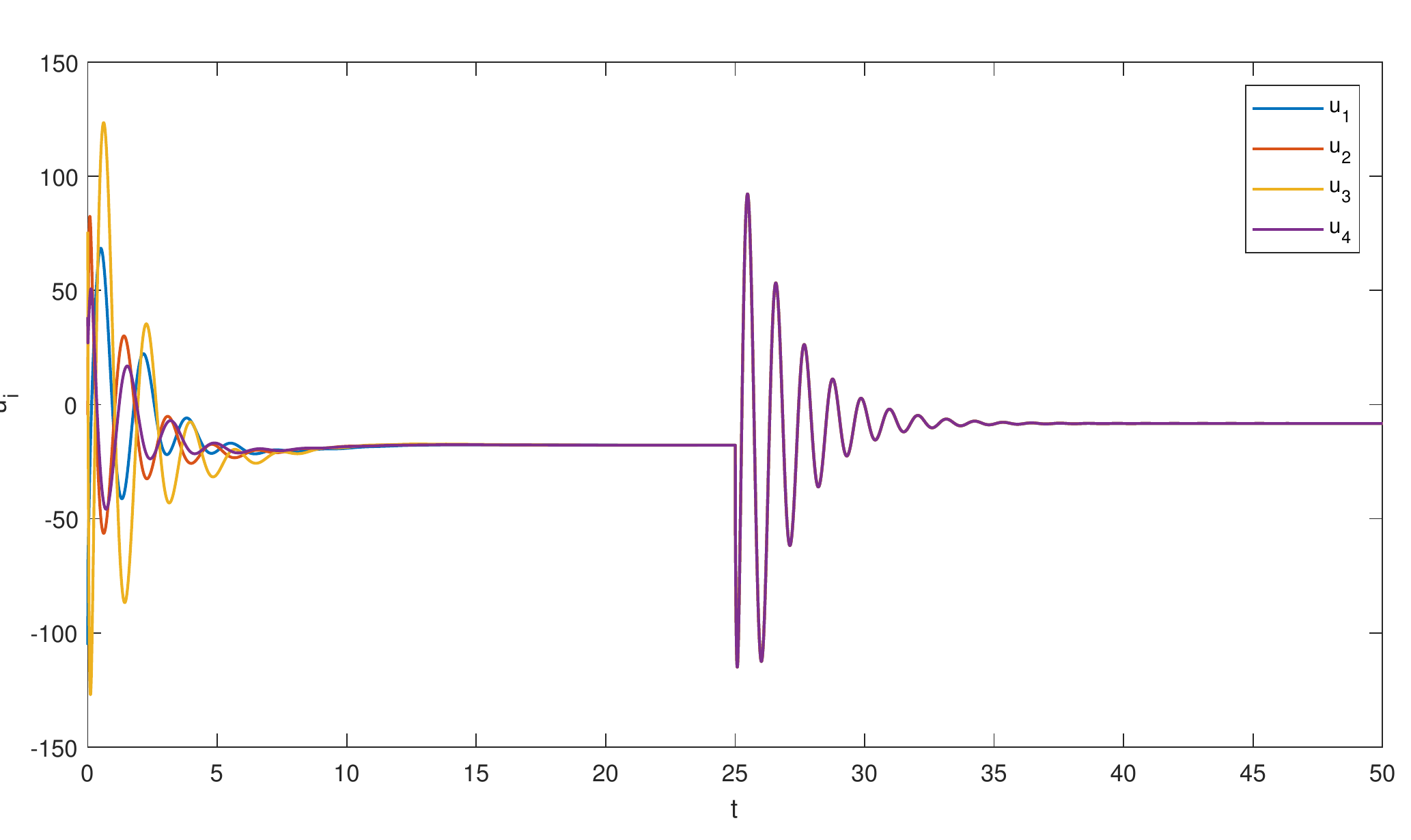}
	\caption{Profile of control efforts under the output feedback controller \eqref{ctr:output}  in {\it Example 2}.}	\label{fig:simu2}
\end{figure}

In the simulation, we choose $k_1=1$, $k_2=2$, $\alpha=1$, $\beta=15$, $\e=6$, and $\gamma=10$ for the controller \eqref{ctr:output}.  The initial conditions for the closed-loop system are (randomly) chosen.  The performance of our proposed optimal signal generator \eqref{sys:generator} is illustrated in Fig.~\ref{fig:generator}.  To make it more interesting, we use two different sets of system parameters.  At first, vector $w$ is chosen as $[0.4,\, 0.3,\,-0.2,\,-0.4]^\top$ before $t=25{\rm s}$. Then we change it to be $[0.1,\, -0.2,\,-0.3,\,0.2]^\top$. With the developed control \eqref{ctr:output}, we present the simulation results in Figs.~\ref{fig:simu1} and \ref{fig:simu2}.  An output consensus on $y^*=3.24$ for these agents is reached even the parameter vector $w$ changes. Meanwhile,  the control signals are found to be bounded and converge to each individual steady-state control effort quickly during these two different phases as reported in Fig.~\ref{fig:simu2}. The simulation results verify the effectiveness and robustness of our distributed output feedback optimal consensus controllers against parametric uncertainties.


\section{Conclusion}\label{sec:con}
We have addressed a robust DOOC problem for a group of uncertain high-order multi-agent systems by measurement output feedback. By inserting an applicable optimal signal generator to reproduce the global optimal point, we have developed a novel distributed output feedback integral controller to solve the problem irrespective of unknown parameters. Future works will include extensions to the cases when this problem has output constraints with more general graphs and external disturbances.

\section*{Acknowledgment}

This work was supported by National Natural Science Foundation of China under Grant 61973043.

%\ifCLASSOPTIONcaptionsoff
%  \newpage
%\fi

\bibliographystyle{ieeetr}
\bibliography{opt-high-abbr}

\end{document}